\documentclass[aps,prl,10pt,a4paper,twocolumn,notitlepage,showpacs,]{revtex4}

\usepackage{amsmath}
\usepackage{times}
\usepackage{amssymb}
\usepackage{graphicx}
\usepackage[unicode=true, pdfusetitle,
 bookmarks=true,bookmarksnumbered=false,bookmarksopen=false,
 breaklinks=false,pdfborder={0 0 1},backref=false,colorlinks=true]
 {hyperref}

\begin{document}

\title{Anderson localization in the quintic nonlinear Schr\"odinger equation}
\author{A. T. Avelar and W. B. Cardoso}
\affiliation{Instituto de F\'isica, Universidade Federal de Goi\'as, 74.001-970, Goi\^ania
- GO, Brazil.}

\begin{abstract}
In the present paper we consider the quintic defocusing nonlinear
Schr\"odinger equation in presence of a disordered random potential
and we analyze the effects of the quintic nonlinearity on the Anderson
localization of the solution. The main result shows that Anderson localization 
requires a cutoff on the value of the parameter which controls the quintic nonlinearity,
with the cutoff depending on the amplitude of the random potential.
\end{abstract}


\maketitle
\emph{Introduction} - The Anderson localization (AL) phenomenon is the suppression of transport due to a destructive interference of the many paths associated with coherent multiple scattering from the modulations of a disordered potential \cite{AndersonPR58}. This effect has been experimentally observed for light in diffusive media \cite{WiersmaNAT97,StorzerPRL06}, photonic crystals \cite{SchwartzNAT07,LahiniPRL08}, optical fiber arrays \cite{SrinivasanPRA08}, microwaves \cite{ChabanovNAT00}, sound waves \cite{HuNP08}, and others. Recently, AL was also observed in noninteracting Bose-Einstein condensate (BEC) \cite{BillyNAT08,RoatiNAT08}. For instance, in Ref. \cite{BillyNAT08} the authors used a $^{87}$Rb BEC in a one-dimentional (1D) waveguide in the presence of a controlled disordered potential generated by a laser speckle for observation of an exponential tail of the spatial density distribution, which is a signature of the AL; the AL was also observed in \cite{RoatiNAT08} for a $^{39}$K BEC in presence of a 1D quasiperiodic bichromatic optical lattice. Motivated by these experimental investigations, many theoretical studies has been done in the last years considering random potentials \cite{SP_PRL07, SP_NJP08, NattermannPRL08, PikovskyPRL08, ChengPRE10} and bichromatic optical lattice \cite{BiOL}.

In BECs, the search for localized structures such as solitons and breathers has also attracted attention of many researchers. As example, we have the experimental observation of dark solitons \cite{dark}, formed in BEC with repulsive $^{87}$Rb atoms, and bright solitons \cite{bright}, generated in attractive $^{7}$Li atoms. From the theoretical point of view, the control of such solutions can be facilitated through the search for analytical solutions of the 1D Gross-Pitaevskii equation (GPE). In this sense, recently, analytical solitonic solutions to the more general case, employing space- and time-dependent coefficients, was considered for the cubic \cite{BB_PRL08}, the cubic-quintic \cite{AvelarPRE09}, the quintic \cite{BB_PLA09}, and also the GPE in higher dimensions \cite{PG_PD06}. Analytical breather solutions has been found in Ref. \cite{breathers}. 

A current challenging problem in BECs is to understand how the nonlinear effects may affect the localization which appears from the presence of random potential. In this sense, the destruction of AL by a weak nonlinearity was recently studied considering a one-dimensional discrete nonlinear Schr\"odinger lattice with disorder \cite{PikovskyPRL08}. In Ref. \cite{PikovskyPRL08}, the authors demonstrated numerically that above a certain critical strength of cubic nonlinearity the AL is destroyed and an unlimited subdiffusive spreading of the field along the lattice occurs. Here, we consider a quintic nonlinear Schr\"odinger equation (QNLSE) in presence of a weak random amplitude potential. Our goal is to study the effects of the quintic repulsive (defocussing) nonlinearity on the AL of the system. In the BEC case, the quintic nonlinearity is related to the three-body scattering \cite{Pethick02}, or to a mere expansion of the cubic term due to a reduction of dimensionality \cite{SalasnichPRA02}, while in nonlinear crystals it is due to the quintic order nonlinear electric susceptibility $\chi^{(5)}$ \cite{Kivshar03}. Recently, we considered the unidimensional reduction of a three-dimensional BEC with two- and three-body interactions \cite{CardosoPRE11}, and there it was shown that the quintic GPE is the effective equation that describes the profile of the BEC when it presents a weak coupling regime, with the cubic nonlinearity being much smaller than the quintic one.

\textit{Theoretical model} - We deal with a 1D quintic NLSE written in its dimensionless form \cite{CardosoPRE11}
\begin{equation}
i\psi_{t}=-\frac{1}{2}\psi_{xx}+V\psi+g_5|\psi|^{4}\psi,\label{eq:1}
\end{equation}
where $\psi=\psi(x,t)$ is the wave-function describing the collective
state of the atoms in a BEC or the electric field propagating in a
nonlinear crystal, $\psi_{t}\equiv\partial\psi/\partial t$
and $\psi_{xx}\equiv\partial^{2}\psi/\partial x^{2}$, $V=V(x)$ is
the potential, and $g_5$ is the nonlinear constant which controls the quintic nonlinearity. 

In the present paper we work with random potentials that are generated by an algorithm 
of random number of the \texttt{MAPLE 13} program called \texttt{RAND}. Here, we use such algorithm
to obtain a pattern of potential that present a random behavior, given by
\begin{eqnarray}
V(x) = V_{0}\xi(x),\label{eq:3}
\end{eqnarray}
with $\xi(x)=A_n$ for $x\in[-L+2nL/S,-L+2(n+1)L/S]$ and $n=0,1,...,S-1$, where $A_n$ is a random number, $2L$ the dimension of the system in spatial $x$ direction and $S$ the number of disordered amplitude points. Also, $V_{0}$ is an amplitude parameter of the random potential in (\ref{eq:3}). We stress that the Eq. (\ref{eq:3}) consists of a perturbation in the amplitude, in the spatial range [$-L,L$].

\textit{Numerical results} - The Eq. (\ref{eq:1}) is numerically solved using a split-step algorithm
based on the Crank-Nicolson method. Here
we use space and time steps given by 0.04 and 0.001, respectively.
First, we employ an imaginary-time method for propagation of a Gaussian
pulse, to get the profile of the ground state solution. Next we realize
various analyzes for the conditions for occurrence of localized solutions in the system. The interval of the system used in the simulations is $[-L,L]$ with $L=30$. We also selected $S=300$,
that is a typical number of perturbed points for the considered range \cite{ChengPRE10}.

\begin{figure}
\includegraphics[width=7cm]{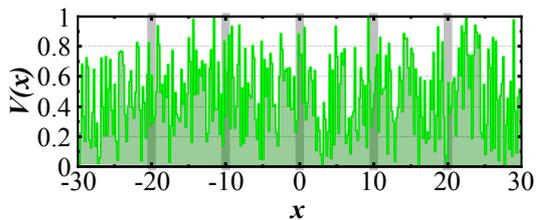}
\caption{(Color online) Random amplitude potential given by Eq. (\ref{eq:3}), for $V_0=1$.}
\label{F1}
\end{figure}

In Fig. \ref{F1} we display the potential with random amplitudes given by Eq. (\ref{eq:3}). In BECs the random potentials can be modeled by laser speckles \cite{LyePRL95,BillyNAT08} that is formed when a laser beam crosses a diffusive plate, and focused onto the atomic sample, creating a random phase shift along the beam profile, which is converted into a random intensity distribution. Also, the disorder can be produced by quasiperiodic lattices \cite{RoatiNAT08} that is generated through combination of a deep primary lattice  with a secondary one via the dipole potential of a laser beam arranged in standing wave configuration. This quasiperiodic lattices also appears in quasicrystals \cite{ShechtmanPRL84}. 

\begin{figure}
\includegraphics[width=7cm]{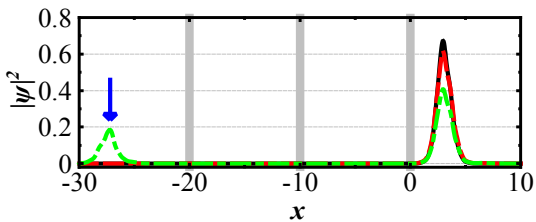}
\caption{(Color online) Profile $|\psi|^2$ of the QNLSE in the presence of a potential with random amplitude considering $g_5=0$ in solid (black) line, $g_5=1$ in dashed (red) line, and $g_5=3$ in dash-dotted (green) line. Also, we have used the values $V_0=5$ and $S=300$. The arrow (blue) shows the position in which the fragmentation starts.}
\label{F2}
\end{figure}

Next we use the simulations to verify the influence of the quintic term of the QNLSE on the localization. Fig. \ref{F2} shows the stationary state obtained via imaginary time evolution of the QNLSE for three distinct values of the nonlinearity. The arrow (blue in online version) shows the position of a second peak associated to the fragmentation of the solution. This result will be clarified below.

\begin{figure}
\includegraphics[width=7cm]{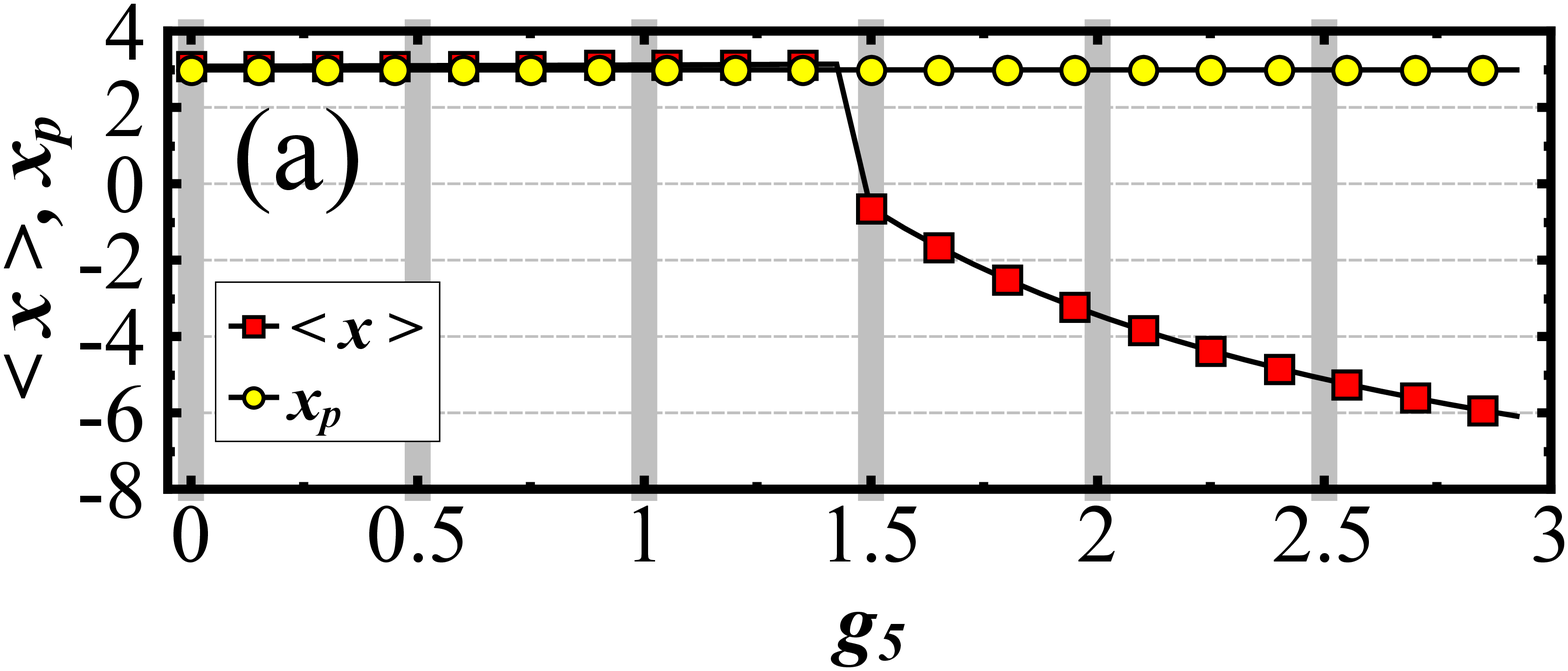}\hfil%
\includegraphics[width=7cm]{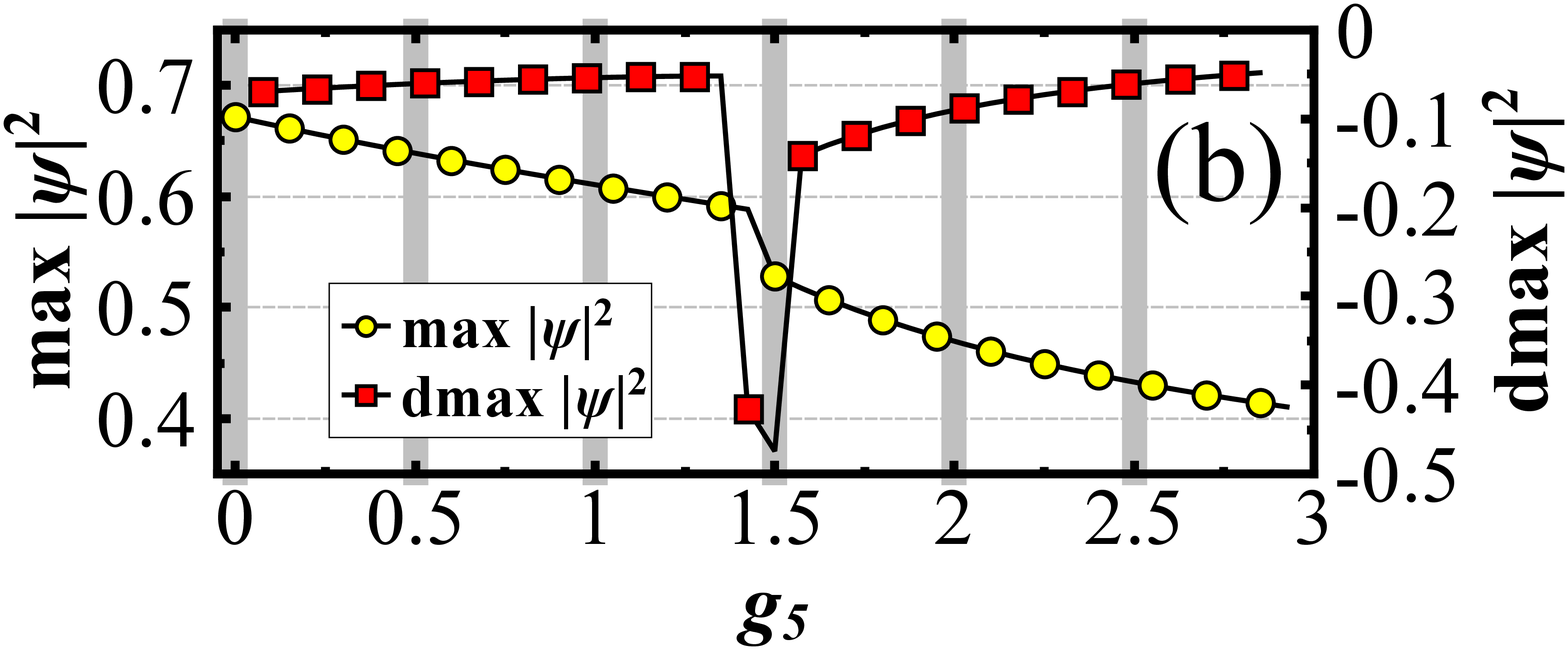}
\caption{(Color online) Influence of the nonlinear quintic term on (a) the average $\langle x \rangle$ (circles) and peak $x_p$ (boxes) positions and (b) the peak height (max $|\psi|^2$, circles) and its derivative (dmax $|\psi|^2$, boxes). A lack of coincidence in the values plotted in (a) indicates a fragmentation in a multi-peak solution. We have used the values $V_0=5$ and $S=300$.}
\label{F3}
\end{figure}

\begin{figure}
\includegraphics[width=7cm]{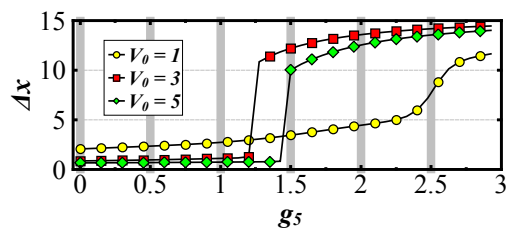}
\caption{(Color online) Root-mean-square width of the profile ($\Delta x$) versus the nonlinear coefficient $g_5$. We display the case with $V_0=1$ in circles (yellow), $V_0=3$ in boxes (red), and $V_0=5$ in diamonds (green).}
\label{F4}
\end{figure}

We have analyzed the average $\langle x \rangle$ and peak $x_p$ positions of the solutions as well as the  peak height (max $|\psi|^2$) and its derivative (dmax $|\psi|^2$) versus $g_5$ as a tentative of verify the influence of the nonlinear quintic term. The results are plotted in Fig. \ref{F3}a and \ref{F3}b, considering $V_0=5$. We look the lack of coincidence in this values as indicative of the fragmentation in a multi-peak solution. Also, in Fig. \ref{F4} we exhibit the root-mean-square  width $\Delta x$ of the profile versus the nonlinear coefficient $g_5$ for different values of the amplitude of the random potential. Note that an abrupt change in $\Delta x$ occurs for a given value of the nonlinearity. This result is more visible for large values of the amplitude of the random potential. For example, considering the case of $V_0=3$ (boxes) the abrupt transition appears at $g_5\simeq 1.2$, and we hope to obtain exponential tails for values of $g_5<1.2$.

\begin{figure}
\includegraphics[width=7cm]{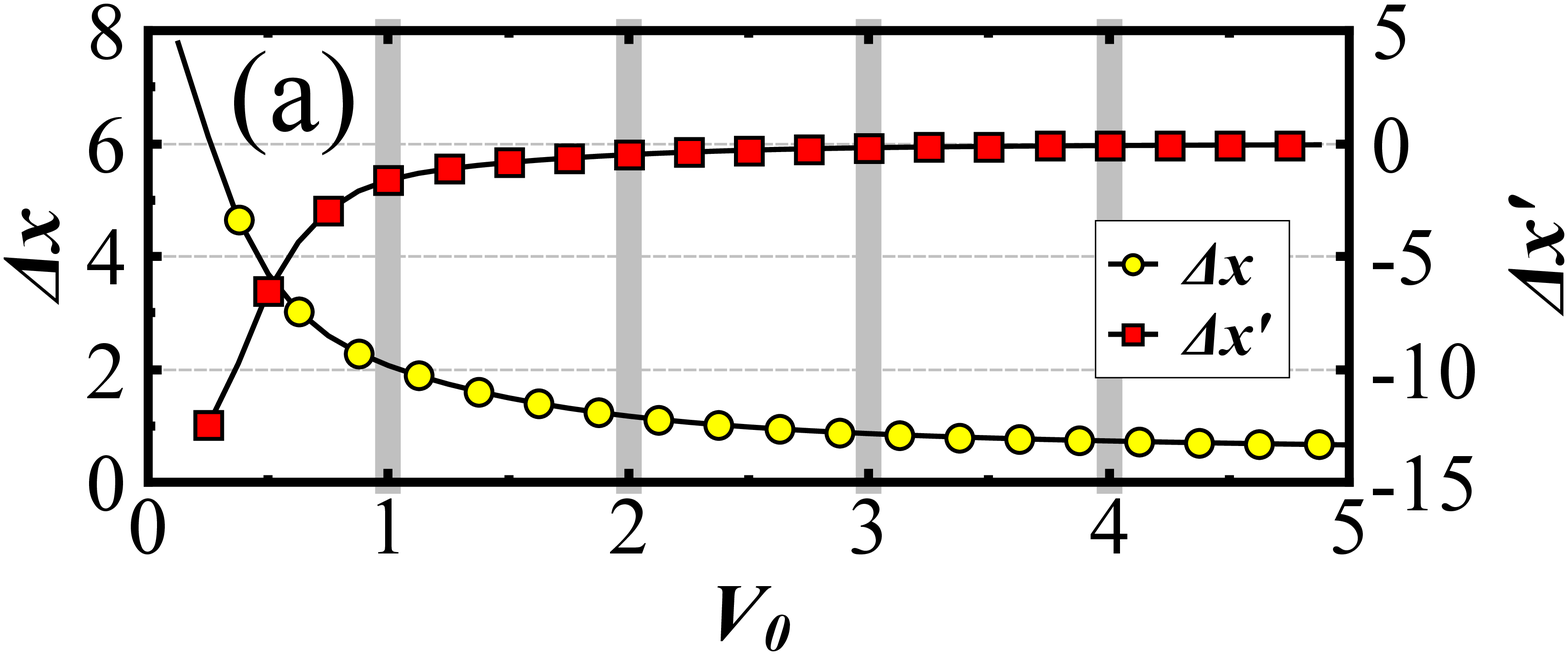}\hfil%
\includegraphics[width=7cm]{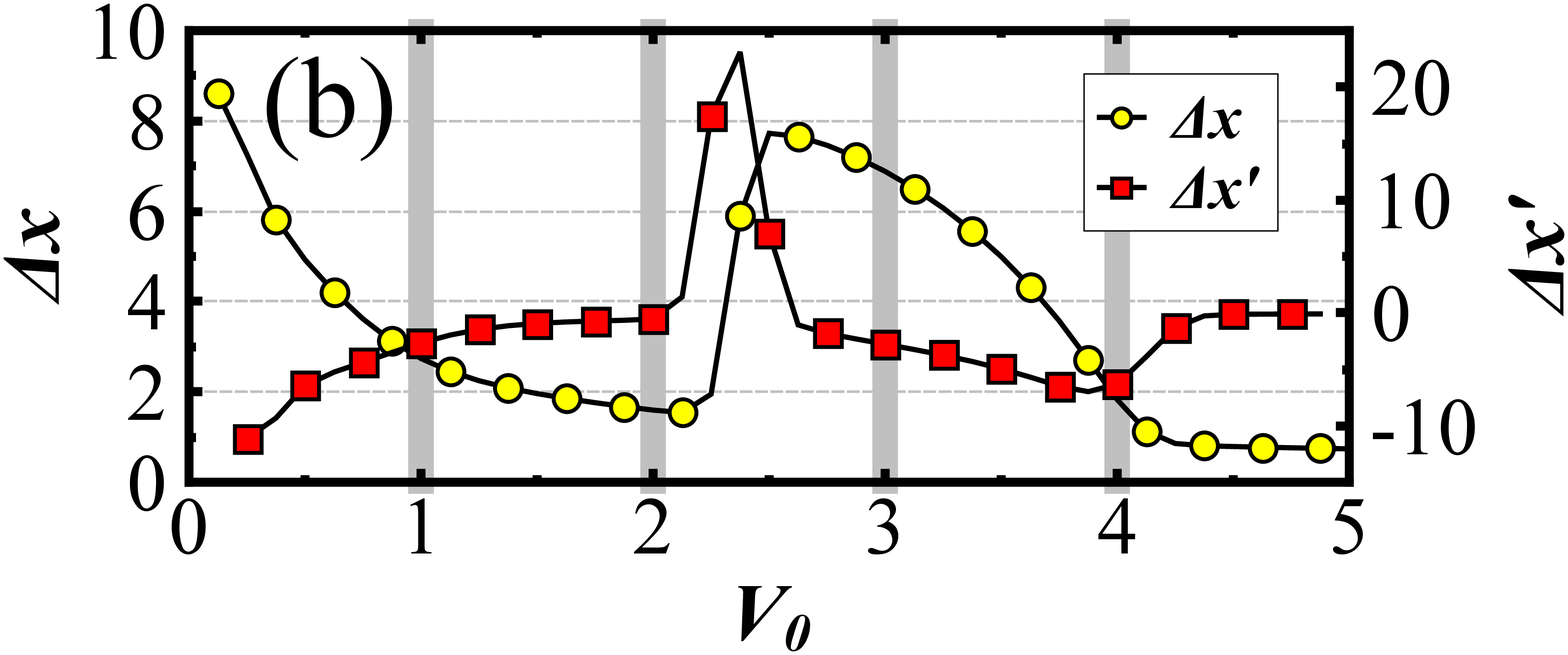}
\caption{(Color online) Root-mean-square width of the profile ($\Delta x$, circles) and its derivative ($\Delta x^{\prime}$, boxes) versus the potential amplitude $V_0$. In (a) we display the case with $g_5=0$ while in (b) it is shown the case with $g_5=1$.}
\label{F5}
\end{figure}

The influence of the amplitude of the random potential on $\Delta x$ and its derivative is displayed in Figs. \ref{F5}a and \ref{F5}b, considering $g_5=0$ and $g_5=1$, respectively. Note that for the case with $g_5=0$, $\Delta x$ is decreased when $V_0$ is increased. This result is expected since for large values of the amplitude of the potential the state tends to become strongly localized \cite{ChengPRE10}. On the other hand, in the presence of a quintic nonlinearity $\Delta x$ suffers abrupt changes. Then, only for a `large' amplitude $V_0$ the state turns localized ($V_0\gtrsim 5$).

\begin{figure}
\includegraphics[width=7cm]{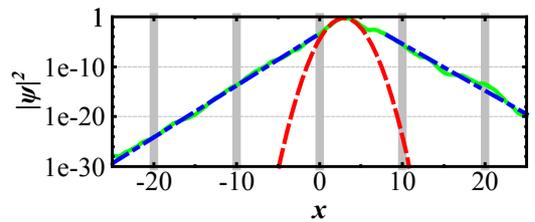}
\caption{(Color online) Example of a profile with AL solution, 
considering $V_0=5$ and $g_5=1$. In solid (green) line we exhibit the profile $|\psi|^2$ of the solution while in dash-dot (blue) line it is shown the exponential fit ($\sim\exp(-2|x-x_p|/l)$). For comparison, we display a Gaussian fit in dashed (red) line.}
\label{F6}
\end{figure}

We have search regions with localized solutions presenting pure exponential decaying tails \cite{BillyNAT08}. Due to the asymmetry of the solution, we employed exponential fit ($|\psi|^2= N \exp [-2|x|/l]$, where $N$ is the amplitude) for the left and right tails. The localized state presents a pure Gaussian tail when the system engender a strong random amplitude potential, since in this case $l\rightarrow 0$. However, the exponential tail is obtained and the limit of localization in a weak random potential is attained when $l > \Delta x$ \cite{BillyNAT08,RoatiNAT08}. In this sense, in Table \ref{T1} we display comparative values of $\Delta x$, left tail $l_L$, and right tail $l_R$ versus $V_0$ considering two specific values for the nonlinearity ($g_5 = 0$ and $g_5 = 1.0$). In the same way, Fig. \ref{F6} shows an example of a profile with AL solution (solid line), the Gaussian (dashed), and exponential tail (dash-dotted) fits, considering $V_0=5$ and $g_5=1$.

\begin{table}
\begin{tabular}{||c|c|c|c||c|c|c|c||}
\hline \multicolumn{4}{||c||}{$g_5 = 0$} & \multicolumn{4}{|c||}{$g_5 = 1.0$} \\
\hline $V_0$ & $\Delta x$ & $l_L$ & $l_R$ & $V_0$ & $\Delta x$ & $l_L$ & $l_R$ \\ 
\hline 0.1 & 8.203 & 7.753 & 5.557 & 0.1 & 8,961 & 8,820  & 5,341 \\ 
\hline 0.2 & 6.532 & 9.416 & 4.507 & 0.2 & 7,530 & 10,558 & 4,705 \\ 
\hline 0.5 & 3.629 & 4.439 & 4.828 & 0.5 & 4,729 & 6,743  & 6,401 \\ 
\hline 1   & 2.066 & 2.306 & 2.764 & 1   & 2,699 & 3,279  & 3,893 \\ 
\hline 2   & 1.170 & 1.441 & 1.471 & 2   & 1,596 & 1,671  & 2,045 \\ 
\hline 5   & 0.660 & 0.717 & 0.800 & 5   & 0,739 & 0,836  & 0,955 \\  
\hline 10  & 0.514 & 0.457 & 0.519 & 10  & 0,553 & 0,526  & 0,547 \\ 
\hline 20  & 0.404 & 0.310 & 0.331 & 20  & 0,443 & 0,336  & 0,358 \\ 
\hline 
\end{tabular} 
\caption{Comparative values of the root mean square width $\Delta x$, left tail $l_L$, and right tail $l_R$ versus the potential amplitude $V_0$ of the potential with random amplitudes considering two values of nonlinearity ($g_5 = 0$ and $g_5 = 1.0$).}
\label{T1}
\end{table}

\begin{figure}
\includegraphics[width=7cm]{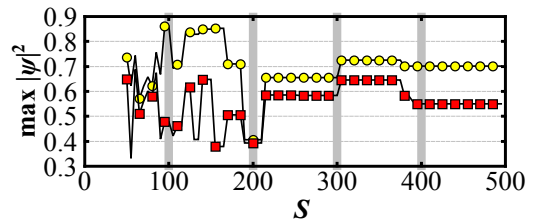}
\caption{(Color online) Peak height (max $|\psi|^2$) vs. the number of
disturbed points S, considering $V_0 = 1$ with $g_5 = 0$ (circles, yellow)
and $g_5 = 1$ (boxes, red).}
\label{F7}
\end{figure}

Finally, in Fig. \ref{F7} we display a simulation of the peak height (max $|\psi|^2$) versus the number of
disturbed points $S$, considering $V_0 = 1$ with $g_5 = 0$ (circles) and $g_5 = 1$ (boxes). Note that in the region $S<200$ the peak height oscillate and for $S>200$ this value is `stabilized'. So, we justify the use of the $S=300$ disturbed points for the potential considered in the present simulations such that a weak change in this value does not interfere in the results. To this end, we have tested the accuracy of the numerical simulation by varying the pattern of the random amplitude potential in Eq. (\ref{eq:3}) and the total number of time and space steps.

\emph{Conclusion} - In conclusion, in the present paper we have consider the quintic defocusing nonlinear
Schr\"odinger equation in presence of a disordered random potential and we have analyzed the effects of the quintic nonlinearity on the Anderson localization of the solution. A cutoff value for the quintic term is obtained related with the amplitude of the random potential, as well as that to the nonlinear Schr\"odinger equation with cubic nonlinearity. The method is general and can be applied for other configurations of disordered potentials.

\subsection*{Acknowledgements}

We would like to thank Dionisio Bazeia for discussions, and for reading the manuscript.
This work was supported by CNPq and FUNAPE-GO, Brazilian agencies.


\end{document}